\documentclass[preprint]{iucr}
\journalcode{M}
\usepackage{graphicx}
\usepackage{physics}
\usepackage{xcolor}
\usepackage{dsfont}
\usepackage{amssymb}

\usepackage{color,soul}
\definecolor{lightblue}{rgb}{0.90,0.95,1}

\definecolor{dred}{rgb}{0.99,0.8,.99}

\let\xxxhat\hat
\renewcommand{\hat}[1]{{\boldsymbol {\xxxhat {#1}} }}
\renewcommand{\vec}[1]{\boldsymbol {#1}}

\begin{document}

\title{A Metropolis Monte Carlo Algorithm for Merging Single-Particle Diffraction Intensities}
\date{2020-08-25}
\cauthor{B. R.}{Mobley}{bemo1722@colorado.edu}{}
\author{K. E.}{Schmidt}{}
\author{J. P. J.}{Chen}{}
\author{R. A.}{Kirian}{}
\aff{Department of Physics, Arizona State University, Tempe, AZ 85287
\country{USA}}

\maketitle


\begin{abstract}
Single-particle imaging with X-ray free-electron lasers depends crucially on 
algorithms that merge large numbers of weak diffraction patterns despite missing 
measurements of parameters such as particle orientations. The 
Expand-Maximize-Compress (EMC) algorithm is highly effective at merging 
single-particle diffraction patterns with missing orientation values, but 
most implementations exhaustively sample the space of missing parameters and may 
become computationally prohibitive as the number of degrees of freedom extend 
beyond orientation angles. Here we describe how the EMC algorithm can be 
modified to employ Metropolis Monte Carlo sampling rather than grid sampling, 
which may be favorable for cases with more than three missing parameters.  Using 
simulated data, this variant is compared to the standard EMC algorithm. Higher 
dimensional cases of mixed target species and variable x-ray fluence are also
explored.  
\end{abstract}

\section{Introduction}

Single-Particle Imaging (SPI) with X-ray Free-Electron Lasers (XFELs) is a 
technique based on the diffraction-before-destruction principle in which 
high-flux, few-femtosecond-duration x-ray pulses produce snapshot diffraction 
patterns before the target is destroyed by the x-rays \cite{Neutze2000, 
Chapman2019}.  The technique is being developed 
because it may enable dynamic macromolecules such as proteins to be imaged at 
atomic spatial resolution and with ultrafast time resolution, without the need 
to cryogenically freeze or crystallize the targets.  The development of SPI has 
been a longstanding challenge because the number of scattered photons from 
isolated molecules is very small, and no practical schemes have emerged that 
can physically orient isolated macromolecules.  Therefore, one cannot calculate 
a simple average of the measured diffraction patterns, and the technique currently
depends crucially on sophisticated data analysis algorithms along with 
instrumentation that eliminates background signals and detector artifacts.

A highly effective solution to the orientation problem was developed by \cite{Loh_2009}, which is referred to as the Expand-Maximize-Compress (EMC) 
algorithm \cite{loh2014minimal, ayyer2016dragonfly}.  EMC uniformly samples the particle orientation space and performs 
likelihood calculations at each orientation sample to measure how well each 2D diffraction pattern 
fits into the 3D model of the particle's Fourier magnitude. The algorithm 
then combines all the diffraction patterns to form a new model, which becomes 
the reference for the next round of orientational likelihood calculations. By 
iterating in this way, the patterns are assembled such that they are 
increasingly consistent with one another, eventually resulting in a locally optimal 
orientation probability distribution for all of the patterns, and a 3D model of the Fourier intensity that is likely to produce the diffraction data.

The EMC algorithm has proven to be highly effective and computationally 
manageable when addressing data for which orientations are unknown. It is 
readily adapted to cases in which more degrees of freedom are needed to model 
the data \cite{Ayyer2021}, but the regular sampling of parameter space that is 
usually employed may become computationally prohibitive.  Such cases may arise, 
for example, when particles are dynamic and cannot be assumed to be exactly 
identical \cite{SchwanderFungOurmazd_2014, Ayyer2021}, or when there are missing parameters 
needed to describe the x-ray source such as time delay or the non-uniform phase 
gradient that is particularly relevant to sub-micrometer focal spots \cite{Gibson_Treacy_2008} \cite{Loh:13}. 
Another situation of interest is the extension of EMC to the weak 
diffraction patterns from crystals \cite{Lanro5012} that are 
expected from compact femtosecond x-ray sources such as the Compact X-ray Light Source (CXLS) currently being developed at Arizona State University \cite{Graves_2017}.  
The use of a uniform grid of orientational samples for crystal diffraction poses 
a particular challenge because the angular extent of Bragg peaks is very small, 
and hence the number of orientation samples will need to be orders of magnitude larger than that for the 
uncrystallized molecule. Diffraction patterns from crystals moreover may require additional parameters to model effects arising from various types of disorder and crystals shapes \cite{Ayyer_etal_2016, Morgan_etal_2019, Chen_etal_2019}.

In many cases, the likelihood of a particular pattern being produced by a given 
set of parameters is effectively zero for an overwhelming majority of points 
in the parameter space.  Thus a subset of the parameter space may produce adequate 
results but with far less computational resources or time.  Monte Carlo 
techniques are an attractive means to reduce the number of samples without prior 
knowledge of which subset to choose.  Such techniques readily fit within the EMC 
framework because the calculating and summing of likelihoods amounts to a numerical integration of a probability density over that space, 
which is commonly handled via Monte Carlo techniques.  Moreover, it is known 
that in the case of numerical integration over 4 or more dimensions, Monte Carlo 
techniques tend to be more computationally efficient when compared against 
uniform sampling approaches.   In the following sections, we describe how a 
Metropolis walk in an augmented orientation space may be employed within the EMC 
framework.

It is noteworthy that, for cases in which a continuum of particle states are 
present and a large number of dimensions or samples are needed, graph-theoretic 
approaches \cite{Fung_etal_2009} are in development that are 
particularly powerful in situations where the unknown model parameters cannot be 
identified prior to the data collection.  Such methods can also be more 
computationally efficient than those based on maximum likelihood, but the model 
manifolds that are produced must be further analyzed in order to relate the 
coordinates to physical properties of the target.  The EMC algorithm differs 
fundamentally in that it enforces a physical model at every iteration.    

In this paper, Section 2 will introduce the mathematical theory behind the standard EMC and the Monte Carlo EMC algorithms. Section 3 describes the implementation details of the Monte Carlo EMC algorithm. Section 4 gives details of the simulations conducted to test the algorithm, with the results presented in Section 5.

\section{Theory}
    
\subsection{Diffraction}

The goal of single-particle diffraction experiments is to reconstruct the 
electron density of target molecules by analyzing the diffracted intensity from 
many diffraction patterns. The electron density of the molecule can be obtained 
by a phase retrieval algorithm if one first has the amplitude of its Fourier 
transform, $|F( \vec{q} )|$, which, under the first-order Born approximation, is 
related to the diffraction patterns by
\begin{equation}
I_j = J_0  \, s_j\, |F( \vec{q}_j )|^2
\label{diffraction_intensity}
\end{equation}
where $I_j$ is the intensity (photon counts) scattered into the $j^{th}$ detector pixel 
at position $\vec{q}_j$ in reciprocal space, $J_0$ represents the incident 
fluence (photons per area), and $s_j$ stands for the product of the 
Thompson scattering cross section, the solid angle of the pixel, the 
polarization factor, and any other multiplicative factors such as detector 
gain at the $j^{th}$ pixel.  The reciprocal-space vectors are related to the incident and scattered 
wavevectors, $\vec{k}$ and $\vec{k}'$ respectively, by 
$\vec{q}=\vec{k}'-\vec{k}$.

With equation (\ref{diffraction_intensity}), we can see one detector pixel 
reading gives us information about one point of $|F(\vec{q})|$ in reciprocal 
space. All the pixels in a single detector readout will give us information 
about points constrained to a sphere by the conservation of energy of the 
incoming and outgoing photons. This sphere is known as the Ewald sphere. In an 
ideal world where we could continuously illuminate the target without radiation 
damage, we could sweep this portion of an Ewald sphere through reciprocal space 
by simply rotating the target, and thus reconstruct $|F(\vec{q})|$ at all points 
in reciprocal space up to some maximum resolution $2\pi/q_\textrm{max}$. 
Instead, we live in a world where, to get any useful signal, we need to use a 
pulse of light so strong that it will instantly obliterate the target molecule, 
giving us weak, shot-noise-dominated diffraction patterns from particles in 
random, unknown orientations. 

The goal of the following algorithms is to use a set of weak, 2D diffraction 
patterns from particles in random orientations, denoted $K$, to construct a 3D 
model, $M$, of the modulus squared of the Fourier transform of the molecule, $|F( 
\vec{q}) |^2$. The $k^{th}$ 2D diffraction 
pattern, referred to as $K_k$, has information corresponding to an Ewald slice  
of the 3D model at some orientation. To insert this information into $M$, the 
value of the diffraction pattern's $j^{th}$ detector pixel, $K_{kj}$, is divided 
by $J_0  s_j$ and is added at the appropriate q-vector into the 3D 
model. Inserting all of the detector pixel values from many diffraction 
patterns at their correct orientations and averaging when there are multiple contributions will reconstruct the 3D model of 
$|F(\vec{q})|^2$ up to the maximum detected resolution.  The problem of unknown orientations can be solved with the EMC algorithm as described below.

\subsection{The expand-maximize-compress algorithm}

The original EMC algorithm \cite{Loh_2009} is based on the Expectation 
Maximization algorithm, and solves the unknown-orientation problem by iterating 
three steps (Expand, Maximize, Compress) that update and improve the 3D model, 
$M$, until it is approximately converged.  The model can be initialized in many 
ways (for example, by merging all patterns with randomized orientations). The 
final model ideally has maximal probability of producing the observed data.  The 
EMC steps are detailed in the following sub-sections.



\subsubsection{Expand step}

The expand step refers to the expansion of the 3D model, $M$, into a set of 
oriented Ewald slices we will call $M_{rj}$.  Here, the index $r$ corresponds to a particular 
orientation $\Omega_r$ in the pre-defined orientation set, which are sampled 
approximately uniformly.  The number of orientations in the set is chosen 
according to Shannon's sampling requirements.  The index $j$ corresponds to the 
$j^{th}$ pixel in the detector, which allows for efficient comparisons of each 
model slice to the measured diffraction patterns, $K_{kj}$.

The expansion allows us to assign a likelihood to each of the orientations.
Since the true orientations of the particles are unknown, the likelihood that 
diffraction pattern $K_k$ corresponds to orientation $\Omega_r$ is calculated 
using the current model, $M$. The diffraction patterns are assumed to be 
well-modeled by independent Poisson random variables, and therefore this 
likelihood is given by
\begin{equation}
	P(K_{k}|M\Omega_{r})=\prod_{j}\frac
		{(J_0 \,s_jM_{rj})^{K_{kj}}\exp\left({-J_0 s_jM_{rj}}\right)}
		{K_{kj}!}
	\label{poisson}
\end{equation}
The beam fluence $J_0$ is assumed here to be constant, a condition that is 
relaxed in Section~\ref{sec_VariableFluenceTheory}.

\subsubsection{Maximize step}

The update rule for the new model, $M^{t+1}$, given the current model, $M^t$, at the $t^{th}$ iteration, and 
the set of diffraction patterns, $K$, is derived though Expectation 
Maximization. The iterative Expectation 
Maximization rule finds the values of the model that will 
maximize the full likelihood of the data. To find this rule for our application, we maximize the function $Q(M^{t+1}|M^t)$, defined as 
\begin{align}
	Q( M^{t+1} | M^t )
	&=\sum_k \sum_r P_{rk}^t \log( P( K_k | M^{t+1} \Omega_r ) )
\nonumber
\\
	&=\sum_{j} \sum_{k} \sum_{r}
		P_{rk}^t \bigg( K_{kj} 
		\log(  J_0  \, s_j M^{t+1}_{ rj} ) - J_0 \, s_j M^{t+1}_{ rj} -\log(K_{kj}!) \bigg) \;.
\label{log_likelihood}
\end{align}
Equation~(\ref{log_likelihood}) can be  thought of as a weighted average of the new model diffraction pattern log 
likelihoods, $\log(P(K_k|M^{t+1}\Omega_r))$, averaged over orientation weighted 
by the probability of that orientation determined from the current model, $P_{rk}^t = P(\Omega_r|M^t K_k)$.

The total orientational probability is normalized to unity over the 
full finite list of orientations such that $\sum_r P_{rk}^t = 1$.
Maximizing $Q$ with respect to the values of $M^{t+1}_{rj}$ results in the 
following update rule:
\begin{equation}
	M^{t+1}_{rj}=\frac
		{\sum_{k} P_{rk}^t\frac{ K_{kj} }{J_0 \,  s_j} }
		{\sum_{k} P_{rk}^t}
\label{eq_EMC_traditional_UpdateRule}
\end{equation}
Equation~(\ref{eq_EMC_traditional_UpdateRule}) updates each of the 2D model 
Ewald slices by taking a weighted average of all the measured patterns. 

\subsubsection{Compress step}

The Maximize step does not update the 3D model $M$; it updates the individual 
Ewald slices. Any pair of the updated Ewald slices $M^{t+1}_{rj}$ will have a 
line of intersection where their values should agree, but the update rule in 
Equation~(\ref{eq_EMC_traditional_UpdateRule}) does not ensure such agreement.  
The Compress step handles this discrepancy by inserting the slices into a new 
3D model in a way that averages the values that fall in the same voxel. The new 
model on a 3D grid in reciprocal space, indexed by $i$, is
\begin{equation}
	M^{t+1}_{i}=\frac
		{ \sum_{k} \sum_{r} \,w_{r}P_{rk}^t 
		[R(  \frac{ K_{k} }{s} ,\Omega_{r})]_{i} }
		{J_0 \, \sum_{k}\sum_{r}  \,w_{r}P_{rk}^t [R(\mathds{1},\Omega_{r})]_{i}}  \;,
\label{3D_update}
\end{equation}
where the function $R(X, \Omega_r)$ inserts its first argument, $X$, which has a 
value for each detector pixel, at the corresponding voxels in a 3D array. The 
details of the function $R(X, \Omega_r)$ are given in appendix 
\ref{Appen_A_trilinear_insertion}. The term in the denominator, 
$R(\mathds{1},\Omega_{r})$, inserts an Ewald slice of entirely ones, which has 
the affect of making sure the value at every voxel is the average of all the 
detector values that contributed to it. The fraction $\frac{K_k}{s}$ is to be 
understood as the corresponding entries of each list being divided such that the 
$j^{th}$ entry of the list $\frac{K_k}{s}$ is $\frac{K_{kj}}{s_j}$.
The coefficient $w_{r}$ is a weight due to geometric considerations of tiling 
orientation space, specifically, from angular deficits when the tiles are on the 
edges or vertices of, for example, the underlying 600 cell tiling scheme as used 
in \cite{Loh_2009} and from the projection of this 600 cell onto the unit sphere. 
The full explanation of this orientation tiling method can be found in appendix C
of Loh et al. \cite{Loh_2009}.
Once the Compress step is complete, the EMC algorithm loops back to the Expand step with the new 3D model $M^{t+1}$, and the sequence of steps is repeated until convergence.


\subsection{Monte Carlo EMC}

A Monte Carlo method is any technique employing repeated sampling of random numbers to perform a calculation,
especially one that, in principle, has a deterministic answer.
We will use the case of Monte Carlo integration to motivate the use of Monte Carlo techniques in the EMC calculation.

We can imagine the orientational sums in the numerator and denominator of Equation~(\ref{3D_update})
as a Riemann integral of some function in orientation space.
One of these sums can be written in the general form
\begin{equation}
	\sum_{r} w_{r} P(\Omega_{r})
		f(\Omega_r)
	=\sum_{r} w_{r} \Delta \Omega_r
		 \frac
		{ P(\Omega_{r})}
		{\Delta \Omega_r}
		f(\Omega_r)
	\approx \int d\Omega \,
		p(\Omega)
		f(\Omega)
\label{eq_MC_theory1}
\end{equation}
$P(\Omega)$ is the probability of a particular orientation like our $P_{rk}^t$ above;
$w_r$ is the same geometric weight; and $f(\Omega)$ is some function on orientation space
representing $[R(K_{k} / s,\Omega)]_{i}$ from before.
$ \Delta\Omega_r$ represents the Voronoi volume in orientation space of our orientation tile $\Omega_r$.
If we take the number of orientations to be arbitrarily large,
$ \Delta\Omega_r$ will approach zero as the orientation tiles get closer together,
and $w_r$ will approach one as the tiling approaches perfect uniformity.
We can  think of $p(\Omega) = { P(\Omega_{r})}/{\Delta \Omega_r}$ as a probability density function in orientation space.
The sum given by Equation~(\ref{eq_MC_theory1}) is approximating a continuous integral over orientation space, which can also be approximated using Monte Carlo integration, that is,
an integration of a probability density with some function can be approximated by sampling the probability density many times
and averaging the function evaluated at those samples, i.e.,
\begin{equation}
	\int d\Omega \,p(\Omega)f(\Omega)=\lim_{N \to \infty}\frac{1}{N}\sum_{n=1}^{N} f(\Omega_n) \;,
\end{equation}
where $\Omega_n$ is the $n^{th}$ sample of the probability distribution $p(\Omega)$.

Replacing the orientational Riemann sums in Equation (\ref{3D_update}) with their Monte Carlo counterparts yields a new update rule
\begin{equation}
	M^{t+1}_{i}=
			   \frac
		{\sum_{k} \sum_{n}\,[R( \frac{K_{k}}{s},\Omega_{kn}^t)]_{i}}
		{J_0 \sum_{k}\sum_{n} \,[R_{}(\mathds{1},\Omega_{kn}^t)]_{i}} \;,
\label{3D_update_MC}
\end{equation}
where $\Omega_{kn}^t$ represents the $n^{th}$ sample from our orientational probability density function, $p(\Omega|K_kM^t)$.
The random sampling of $\Omega_{kn}^t$ is achieved, in our case, through a Metropolis walk.
The explicit definition of $p(\Omega|K_kM^t)$
and details of the Metropolis walk will be described in the following sections.

\subsection{Metropolis Sampling}
A Metropolis walk \cite{MRRTT_1953} is a method of sampling a difficult probability distribution, $P(\sigma)$,
that employs a random walk around the space governed by a transition probability, $P_{T}(\sigma \rightarrow \sigma_{new})$,
the probability of transitioning from the current state, $\sigma$, to a new state, $\sigma_{new}$.
The method is often based on the principle of detailed balance, that is,
\begin{equation}
	P(\sigma)P_{T}(\sigma \rightarrow \sigma_{new})=P(\sigma_{new})P_{T}(\sigma_{new} \rightarrow \sigma)
\label{detailed}
\end{equation}
which says the total frequency at which the walk transitions from point A to B will be the same as that of point B to A,
and that, consequently, the walk is in statistical equilibrium with the distribution.
We also need to satisfy that the walk cannot get stuck in subcycles in the space,
and that the walk can cover the whole space in a finite number of steps.
If we can define a transition probability that satisfies these principles,
the resulting walk after enough steps will reflect a set of samples from the desired distribution.
The Metropolis method achieves this by factoring $P_{T}(\sigma \rightarrow \sigma_{new})$
into an easily sampled distribution of our choosing, $g(\sigma \rightarrow \sigma_{new})$, called a proposal distribution, and an appropriate factor called the acceptance ratio, $A(\sigma_{new},\sigma)$. Thus,
 \begin{equation}
	P_{T}(\sigma \rightarrow \sigma_{new})=
	A(\sigma_{new},\sigma)g(\sigma \rightarrow \sigma_{new}) \;. 
	\label{acceptance}
\end{equation}
Combining Equations (\ref{detailed}) and (\ref{acceptance}) yields
\begin{equation}
	\frac
		{P(\sigma_{new})g(\sigma_{new} \rightarrow \sigma)}
		{P(\sigma)g(\sigma \rightarrow \sigma_{new})}
	=\frac
		{A(\sigma_{new},\sigma)}
		{A(\sigma,\sigma_{new})}.
\label{eq_Mwalk_ratio1}
\end{equation}
Equation~(\ref{eq_Mwalk_ratio1}) does not uniquely specify $A$, so we are free to choose
\begin{equation}
	A(\sigma_{new},\sigma)=\mathrm{min}\left(1,\frac
		{P(\sigma_{new})g(\sigma_{new} \rightarrow \sigma)}
		{P(\sigma)g(\sigma \rightarrow \sigma_{new})}\right)
\label{eq_acceptance_ratio1}
\end{equation}
If our proposal distribution is symmetric, i.e., $g(\sigma \rightarrow \sigma_{new})=g(\sigma_{new} \rightarrow \sigma)\,$, Equation~(\ref{eq_acceptance_ratio1}) simplifies to
\begin{equation}
	A(\sigma_{new},\sigma)=\mathrm{min}\left(1,\frac
		{P(\sigma_{new})}
		{P(\sigma)}\right) \;.
	\label{sym_accpt}
\end{equation}
To implement the Metropolis walk, we choose a simple proposal distribution to sample, for example, a normal distribution centered around the current state.
At each step we sample $g(\sigma \rightarrow \sigma_{new})$, then sample a uniform random variable, $u$, between 0 and 1.
If $u$ is less than $A(\sigma_{new},\sigma)$, we accept the proposed step and move from $\sigma$ to $\sigma_{new}$.
If $u$ exceeds $A(\sigma_{new},\sigma)$, we reject it and stay at $\sigma$.
In this way, our simple proposal distribution is pruned into the proper transition probability.
In the next section we will detail how this method can be applied to the problem at hand by taking the state space to be orientation,
and using the walk to approximate samples from an orientational probability distribution.

\subsection{Orientational Metropolis Walk}
\label{Orientational_Metropolis_Walk}

\begin{figure}
	\centering
 	\includegraphics[width=0.5\linewidth]{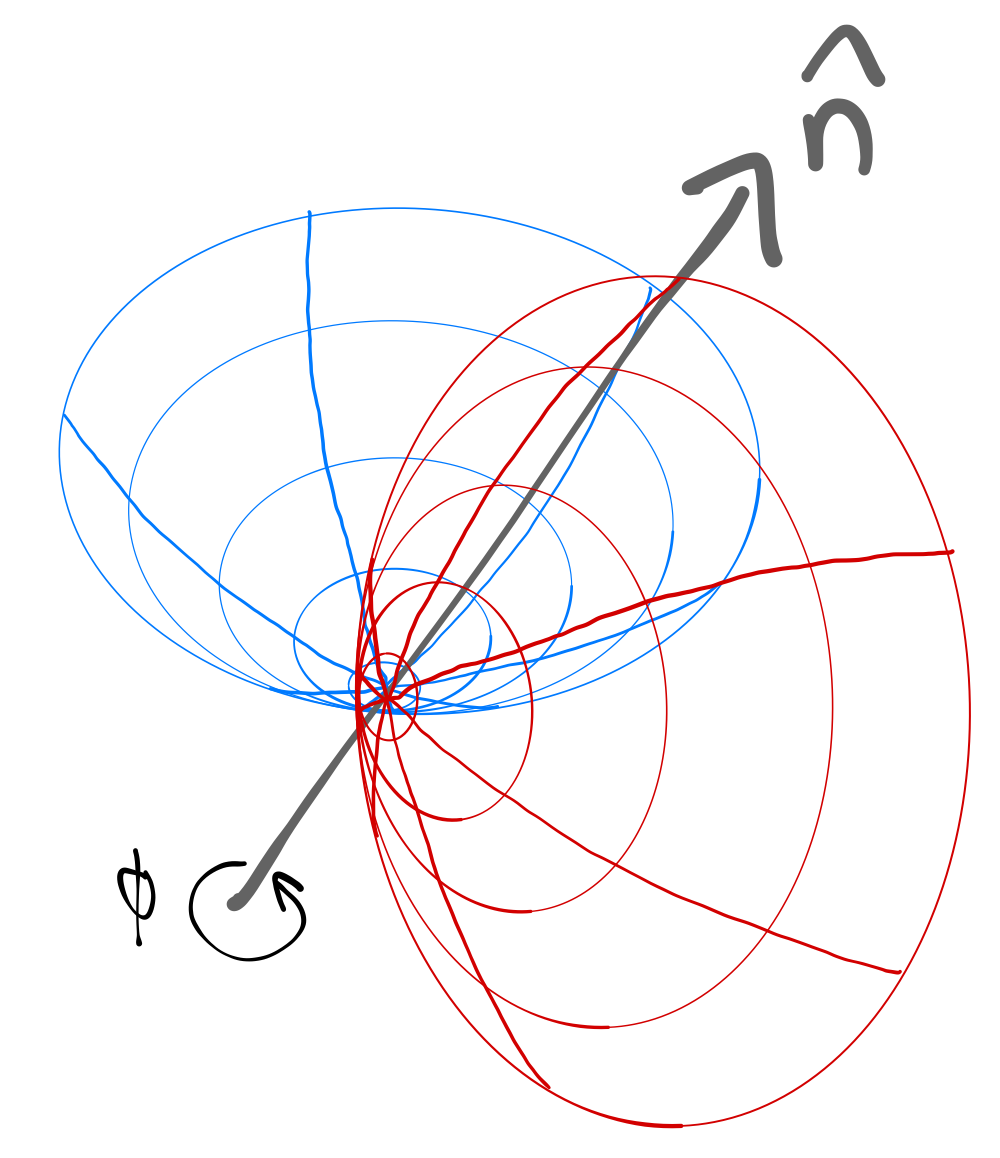}
 	\caption{A section of an Ewald sphere rotated about axis $\hat{n}$ by angle $\phi$. The red shape is being rotated into the blue.}
 	\label{prop}
\end{figure}

In order to sample the orientational probability distributions required for our Monte Carlo model update rule given by Equation~(\ref{3D_update_MC}),
we need to define the corresponding Metropolis walks in orientation space.
First, we need an easily sampled proposal distribution to propose steps in our walk.
Arbitrary orientation can be defined by rotation about a specific axis by a specific angle from a reference orientation,
so the chosen method of proposing steps is to sample a random axis and angle and to perform the rotation they define on the points of an Ewald sphere.
This is implemented by choosing a uniformly random axis and sampling an angle from a Gaussian truncated to the interval from $\pi$ to $-\pi$. 
A cartoon of a section of an Ewald sphere rotated about an axis is shown in Figure \ref{prop}.
Because our proposal distribution is symmetric ( i.e. $g(\Omega \rightarrow \Omega_{new})=g(\Omega_{new} \rightarrow \Omega)\,$),
the exact formula for the proposal distribution is not needed as shown in Equation (\ref{sym_accpt}).

Now that we can propose steps in the walk, we need to define our rules for accepting or rejecting them.
As described in the previous section, this is carried out by drawing a uniform random variable between 0 and 1,
and accepting the step if our draw is less than the value of our acceptance ratio.
The acceptance ratio is a function of our orientational probability density
which we define as follows for the $k^{th}$ diffraction pattern
\begin{equation}
	A_{k}^t(\Omega_{new},\Omega)=
	\mathrm{min}
	\left(1,
	\frac
	{p(\Omega_{new}|K_kM^t)}
	{p(\Omega|K_kM^t)} 
	\right) \;.
\end{equation}
We can calculate $p(\Omega|K_kM^t)$ by using Baye's Theorem to relate this probability density to the likelihood of the $k^{th}$ diffraction pattern,
$P(K_k|\Omega M^t)$.
If we assume a uniform orientational prior, $p(\Omega|M^t)$,
we see that the probability and likelihood are proportional, and proportionality is all we require in the Metropolis acceptance ratio, giving us
\begin{equation}
	p(\Omega|K_kM^t)=
	\frac
	{P(K_k|\Omega M^t) p(\Omega|M^t)}
	{P(K_k|M^t)}
	\propto P(K_k|\Omega M^t) \; .
\end{equation}
To define the likelihood, $P(K_k|M\Omega)$, we imagine all the pixel values of the diffraction pattern, $K_{kj}$,
as independently following Poissonian statistics with mean parameters given
by the 3D model interpolated at the corresponding q-vector position of each pixel.
This can be written as
\begin{equation}
	P(K_k|M^t\Omega)=\prod_{j} \frac{\Big( J_0  \, s_j \, E_{j}(M^t, \Omega)\Big)^{K_{kj}}\exp\Big({ - J_0  \, s_j \, E_{j}(M^t,\Omega)}\Big)}{K_{kj}!}
\label{poisson_all_orientation}
\end{equation}
where $E_{j}(M^t,\Omega)$ is a function that interpolates the 3D density model, $M^t$, on an Ewald sphere in orientation $\Omega$
to find the value at the q-vector of the $j^{th}$ detector pixel.
Equation~(\ref{poisson_all_orientation}) is similar to Equation (\ref{poisson})
but is defined over any orientation instead of the discrete set of orientations from before.
The details of the function $E(M^t,\Omega)$ are given in appendix \ref{Appen_B_trilinear_interp}.

To get good samples of the target probability density, we want the walk to be uncorrelated with the starting position.
To achieve this, the first steps of the walk are disregarded (i.e. not included as part of our sampling) as a ``burn-in period."
The number of steps to be thrown out need to be substantially larger than the autocorrelation time of the steps.

\subsection{Extensions}
The advantage of using of a Metropolis walk is when the dimensionality of the space
causes exhaustive approaches to become too computationally expensive to be feasible.
Extra dimensional spaces could manifest in single particle diffraction experiments
in any additional degrees of freedom that cause the diffraction patterns to differ between shots other than orientation.
Here we present two extensions to the Monte Carlo EMC that both demonstrate performing the walk over a four dimensional space.
The two extension cases are (1) variable fluence, and (2) a two-molecule mixture.

\subsubsection{Variable Fluence}
\label{sec_VariableFluenceTheory}
Here, we will relax our condition that the incident x-ray beam is perfectly consistent. In reality, there is a large amount of variation in beam fluence per shot.
In addition, the cross sectional profile of the beam is not uniform, and it is difficult to control what part of the beam strikes a target molecule.
This forces us to reconsider our diffraction equation (\ref{diffraction_intensity}) as
	\begin{equation}
	I_j = J  s_j\, |F( \vec{q}_j )|^2
	\label{diffraction_intensity_J}
	\end{equation}
where $J$ is unknown and different for every shot. Former algorithms have taken various approaches to this problem of variable fluence \cite{ekeberg20173d}.
We will take it is as another hidden variable in our expectation maximization update rule,
which will simply manifest as another dimension of our Metropolis walk designed to sample the new probability distribution $p(\Omega J|K_kM^t)$.
Our update rule becomes
\begin{equation}
	M^{t+1}_{i} = \frac
		{\sum_{k} \sum_{n}\,[R( \frac{K_{k}}{s},\Omega_{kn}^t)]_{i}}
		{ \sum_{k}\sum_{n} \,J_{kn}^t[R_{}(\mathds{1},\Omega_{kn}^t)]_{i}} \;,
\label{3D_update_MC_J}
\end{equation}
following the notation of Equation (\ref{3D_update_MC}) and 
$J_{kn}^t$ represents the $n^{th}$ fluence sample of $p(\Omega J|K_kM^t)$.
This update rule weights the shots by the incident fluence when averaging contributions,
so the higher fluence shots will contribute more strongly to the final result.
Higher fluence shots have higher photon counts and thus higher information content, so this seems reasonable. 
Changing the expected photon counts changes our likelihood calculation for a particular diffraction pattern.
This gives us
\begin{equation}
	P(K_k|M^t\Omega)=\prod_{j} \frac{ ( J \, s_j \, E_{j}(M^t, \Omega) )^{K_{kj}}\exp\left({ - J  \, s_j \, E_{j}(M^t,\Omega)}\right)}{K_{kj}!}
\end{equation}
following the same notation as Equation~(\ref{poisson}).
The extended probability density we wish to sample, $p(\Omega J|K_kM^t)$, is related to the likelihood according to Bayes' theorem
\begin{equation}
	p(\Omega J|K_kM^t) \propto P(K_{k}|\Omega J M^t)p(J) \;.
\end{equation}
The prior distribution on fluence, $p(J)$, can be constrained if we have a rough estimate of what the x-ray fluence may be.
This knowledge is naively implemented as a uniform distribution inside some interval and zero outside,
which is stating that it is impossible for the fluence to be any value outside the estimated bounds.
This has the affect of giving our walk guide rails that it cannot cross,
which is rather important since the fluence dimension is infinite in extent unlike the compact orientation space.
Implementing a prior prevents the Metropolis walk from meandering arbitrarily far away from a realistic value, where the samples are meaningless.
Though much more sophisticated prior distributions can be imagined based on experimentally determined characteristics of the beam,
the piecewise uniform distribution described has the attractive simplicity of manifesting in the algorithm merely as a rejection
in the case that a step to a fluence value outside the bounds is proposed.

To define our Metropolis walk in the four dimensional orientation plus fluence space, we need to define our step proposal and acceptance ratio.
We will reuse our same orientation proposal from section  \ref{Orientational_Metropolis_Walk}
and combine it with a Gaussian fluence proposal centered around the current fluence.
Both of these proposals are symmetric between the new and old state, so our acceptance ratio will take the simple form
\begin{equation}
	A_{k}(\Omega_{new},J_{new},\Omega,J)=
	\mathrm{min}
	\left(1,
	\frac
	{ p( K_k  |\Omega_{new} J_{new} M ) p( J_{new} ) }
	{ p( K_k | \Omega J M) p( J ) } 
	\right) \;.
\end{equation}
With this, we can generate the fluence and orientation samples necessary for our update rule.

\subsubsection{Two State}
Any additional quantifiable degree of freedom in the target can theoretically be sampled by a Metropolis walk.
One of particular interest is a protein that could take on multiple conformations;
another is imaging interactions between different particles.
The simple case of a two-state target was explored as a proxy for these types of degrees of freedom.
This was implemented by simulating the diffraction data from two different proteins,
which we will refer to as $a$ and $b$, imagined to be injected into the beam together in some ratio.
In this case, once again considering fluence to be constant, we have an additional variable, which we will call $\sigma$,
which stands for which species, $a$ or $b$, a diffraction pattern is considered to have been sampled from.
Our walk will wander between $a$ and $b$, simultaneously constructing two models as it goes. Our update rule for the two state case is 
\begin{equation}
	M^{t+1}_{i\sigma} = \frac{1}{J_0 }
		\frac
		{\sum_{k} \sum_{n} \delta_{\sigma \sigma_{kn}^t}\,
		[ R( \frac{ K_{k} }{s}, \Omega_{kn}^t ) ]_{i} }
		{\sum_{k}\sum_{n}  \, \delta_{\sigma \sigma_{kn}^t} \, [R_{}(\mathds{1},\Omega_{kn}^t)]_{i}} \;,
\label{3D_update_MC_2}
\end{equation}
following the notation of Equation~(\ref{3D_update_MC}), and
where the $n^{th}$ state sample of the joint probability $p(\Omega \sigma|K_kM^t)$ 
is represented by $\sigma_{kn}^t$. $\delta$ is a Kronecker delta that enforces that the diffraction pattern is only added into the model
whose $\sigma$ index matches the $\sigma_{kn}^t$ of that particular draw
(i.e. $\delta_{\sigma \sigma_{kn}^t} = 1$ if $\sigma=\sigma_{kn}^t$ and is otherwise zero).

Our diffraction pattern likelihoods now depend on the state from which we are considering them to have been drawn from, giving
\begin{equation}
	P(K_k|M^t\Omega \sigma) = \prod_{j} \frac
	{( J_0 \, s_j \, E_{j}(M_\sigma^t, \Omega))^{K_{kj}}e^{ - J_0  \, s_j \, E_{j}(M_\sigma^t,\Omega)}}{K_{kj}!} \;.
\end{equation}
Our joint probability distribution of orientation and state, $p(\Omega \sigma|K_kM)$, is
\begin{equation}
	p(\Omega \sigma|K_kM^t) \propto P(K_{k}|M_\sigma^t\Omega_{r})P(\sigma) \;,
\end{equation}
where our prior, $P(\sigma)$, could be estimated as a Bernoulli random variable with some ratio of being $a$ or $b$.
We choose to assume no prior knowledge of the ratio of incoming particles.
With the uniform prior, the Metropolis walk proposes steps that are a random selection of one of the states as well as a random rotation.
As the reconstruction progresses and the two models differentiate, the state proposals will likely be rejected whenever the wrong model is proposed,
and the walk will begin to automatically sort the diffraction patterns into their corresponding models.
The Metropolis walk is defined by the same orientational proposal as in previous sections
and a state proposal that is an evenly distributed Bernoulli random variable regardless of current state.
These are again both symmetrical so we get an acceptance ratio defined as 
\begin{equation}
	A_{k}^t( \Omega_{new}, \sigma_{new}, \Omega, \sigma ) =
	\mathrm{min}
	\left( 1,
	\frac
	{ p( K_k  |\Omega_{new} \sigma_{new} M^t ) }
	{ p( K_k | \Omega \sigma M^t) } 
	\right) \;.
\end{equation}
With this, we can generate the state and orientation samples necessary for our update rule.

\section{Algorithm}

The main structure of the algorithm is a loop that performs the appropriate update rule, given by Equations (\ref{3D_update_MC}), (\ref{3D_update_MC_J}), or (\ref{3D_update_MC_2}),  on the 3D model some preset number of times.
The loop is initialized with a 3D array of noise values drawn from a Poisson distribution whose mean parameter is the mean of all the diffraction pattern values.
This initial model's random anisotropy will determine the orientation of the model reconstruction
and in which box either state is constructed in the case of the two-state algorithm.

The update rule has a sum over the index counting diffraction patterns
which adds the results of a Metropolis sampling for each diffraction pattern based on the same current model.
This structure allows us to distribute the job of performing the Metropolis samplings across multiple cores to be computed in parallel and subsequently combined.
This is achieved by handing each of the cores a copy of the same current model and a different diffraction pattern
and having it calculate the Metropolis sampling of orientation and any additional parameters.
As the cores perform the Metropolis walk, they keep a running sum of the detector pixel values inserted along the oriented Ewald sphere,
which we will call the data sum,
and a separate running sum of an array of ones inserted at the same positions to keep a tally of how many times a voxel has been contributed,
which we will call the tally sum.
The majority of steps are rejected which causes long periods when the walk is stationary,
so, instead of repeatedly adding the same values to the model,
it waits until a step is accepted to add the values multiplied by the length of the pause in the walk in order to save computing time.
When all of the cores have finished their Metropolis sampling and sums,
the data sums from all of the cores are added together voxel-wise and the same for the tally sums.
The combined data sums are then divided by the combined tally sums
which has the result of converting all of the values from the sum of all the contributions at each pixel to the average of those contributions.
The resulting object is our new model and it is then distributed to the cores to start the next iteration.
The algorithm stops after completing the pre-set number iterations of the loop and outputs the models generated by each iteration.

\begin{figure}
	\centering
 	\includegraphics[width=0.8\linewidth]{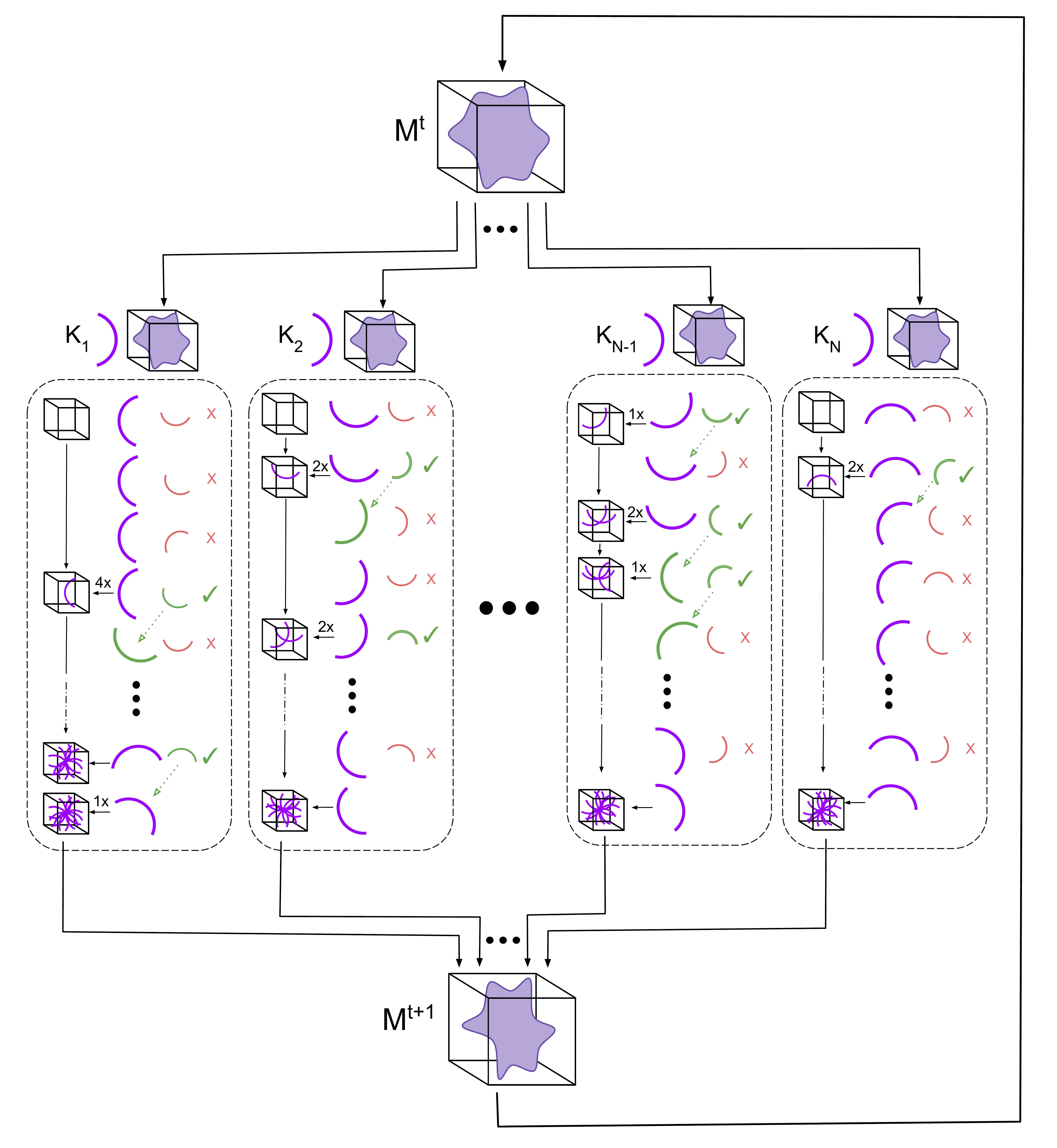}
 	\caption{Pictured is the main loop of the Monte Carlo EMC algorithm.
		The top cube represents the current model labeled $M^t$. Copies of the current model are distributed to cores,
		which are also handed a diffraction pattern labeled by $K_k$.
		This diffraction data corresponds to points on an Ewald sphere, which is pictured as a purple C shape.
		Each core carries out a Metropolis walk dependent on $M^t$ on its diffraction pattern pictured inside the dotted boxes.
		The middle column inside the box represents the orientation of the Ewald sphere at each step of the walk.
		The right column depicts the proposed steps that are accepted ($\checkmark$) or rejected (X).
		The left column depicts values of the diffraction pattern being inserted along the oriented Ewald sphere into a 3D array.
		Once all of the walks have finished, the resulting 3D arrays are combined into the new model, labeled $M^{t+1}$.
		The new model is used as the reference model of the next iteration.}
 \label{Alg}
\end{figure}

\begin{figure}
	\centering
 	\includegraphics[width=0.8\linewidth]{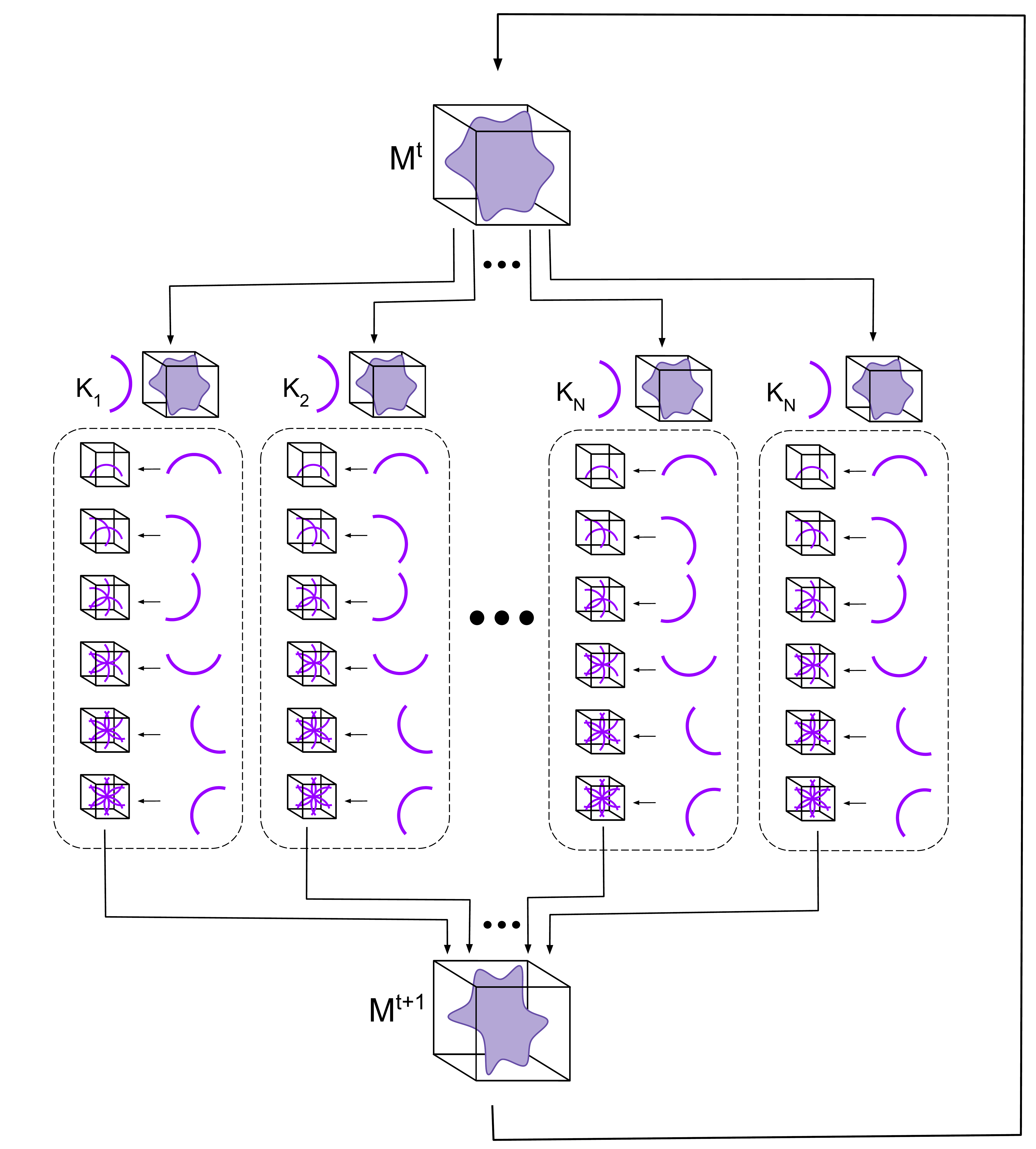}
 	\caption{Pictured is the main loop of the regular sampling EMC algorithm.
		The top cube represents the current model labeled $M^t$.
		Copies of the current model are distributed to cores which are also handed a diffraction pattern labeled by $K_k$.
		This diffraction data corresponds to points on an Ewald sphere, which is pictured as a purple C shape.
		Each core loops through the same list of orientations of the Ewald sphere, schematically represented here by six orientations,
		whereas, in practice, there are on the order of 1000.
		For each orientation, the core calculates a weight dependent on $M^t$ and inserts the values of $K_k$ at that orientation and weight into a 3D array.
		The resulting 3D arrays from all the diffraction patterns are combined into the new model, labeled $M^{t+1}$.
		The new model is used as the reference model of the next iteration.}
 \label{Alg_bf}
\end{figure}

The Monte Carlo EMC algorithm differs from the regular sampling EMC algorithm only in how we calculate the update rule in the Maximization step.
Flow diagrams for the main loop of the Monte Carlo and regular sampling algorithms are shown in Figures \ref{Alg} and \ref{Alg_bf}, respectively.

\section{Simulation}
\subsection{Simulated Data}
\label{sim_data}
The Cytochrome-c (Protein Data bank ID: 3CYT) was selected as the model protein to reconstruct.
Its modulus squared 3D form factor, $|F(\vec{q})|^2$, as described in Equation~(\ref{diffraction_intensity}),
was calculated according to the Born approximation as the square modulus of the Fourier transform of the protein's electron density.

To simulate diffraction patterns, we need to define the geometry of our simulated detector array.
The geometry of the detector ( i.e. distance form the target molecule, pixel size, etc.) determines the angular size of the pixels. 
The detector was simulated with the reborn software package.
The simulated detector is a single panel of 51 pixels by 51 pixels with a pixel size of four mm$^2$ at a distance of half a meter.

\subsubsection{Constant Fluence Data}
\label{sec_constant_fluence_data}

To create a constant fluence data set, we first simulate the photon fluence from a typical XFEL beam
with five millijoule pulse energy, five keV photon energy, and a micron diameter defined by full width at half maximum.
This yields a fluence, $J_0$, of $7.95\times 10^{26}$  photons per meter squared.
With all of our terms defined, we can use Equation~(\ref{diffraction_intensity}) to calculate the fluence incident on any given pixel by interpolating $|F(\vec{q})|^2$ at the pixel's q-vector
and using its geometrically defined solid angle to calculate an expected number of incident photons per shot.
The noise is assumed to be purely Poissonian, so, to get a simulated pixel reading,
a Poissonian random variable is sampled with that pixel's expected fluence as its mean parameter.

The data we want to simulate are suppose to represent exposures of individual molecules at random orientations.
The q-vector of a pixel depends on the position of the pixel on the detector and the orientations of target molecule.
All of the detector pixels' q-vectors together are points on an Ewald sphere at some orientation in reciprocal space.
To simulate a particular shot, we sample a uniformly random orientation, taking note of it for later use,
then simulate the set of pixel values by interpolating at their positions on the Ewald sphere at that orientation.
Four thousand diffraction patterns were generated. Their average photon count per diffraction pattern was about 4400 photons.
The number of initial Metropolis walk steps to be thrown out for samples to become uncorrelated was estimated to be about 150 steps in the constant fluence case and thus a burn-in period of 200 steps was chosen.
 
 \subsubsection{Variable Fluence Data}
 \label{sec_variable_fluence_data}
 
The variable fluence data is simulated in the same way as the constant fluence data except that each shot needs to have a  random fluence.
This is implemented by sampling  a gaussian centered around the mean fluence value, $J_0$, for each shot, and noting it along with the orientation for later use.
The mean fluence is set to be the same value $J_0$ as the constant fluence data, and the standard deviation of our gaussian was set to be one tenth of the mean.
With the random orientation and fluence,
we can use Equation~(\ref{diffraction_intensity_J}) to find the mean parameters corresponding to each pixel
and sample the Poisson random variables to yield our diffraction patterns. A set of four thousand diffraction patterns were generated.
Their average photon count was also about 4400.

 \subsubsection{Two-State Data}
 \label{sec_two_state_data}
  
 To simulate a mixture of particles, we need a second protein with which to simulate diffraction patterns.
 Lysozyme, (Protein Data Bank ID: 2LYZ) was selected for this purpose. Its modulus squared form factor is calculated as in section \ref{sim_data}.

The two-state data is simulated exactly like the constant fluence data, except for each shot it is randomly decided which molecule's form factor to use.
This is decided by a Bernoulli random variable with a ratio of 40\% cytochrome to 60\% Lysozyme.
A set of 4000 diffraction patterns were simulated with an average photon count of about 3300.

\subsection{Measures of Success}

We need an object with which to compare the algorithm's reconstruction to judge its degree of success.
We could use the molecules' modulus squared form factor interpolated on a grid, as this is what we are trying to construct after all,
but, instead, we will use the best possible merge of the data, defined as the data with all hidden parameters known.
This has the advantage of having all same artifacts of interpolating and reinserting
that will appear in the algorithm's output already baked in, thus avoiding something of an apples-to-oranges situation.
The crucial information hidden from the algorithms, orientations, fluence, and states, is used to construct the best possible merges of the data sets.

\subsubsection{Best Possible Constant Fluence Reconstruction}

To construct the best possible merge in the constant fluence case,
we only need to insert the data into a 3D array at the same position from where they were sampled.
Averaging where there are multiple contributions to a voxel and accounting for solid angles and fluence,
we reconstruct the best possible estimate of $|F(\vec{q})|^2$ which we'll call $M^{best}$.
It is defined as
\begin{equation}
	M^{best}_{i}=\frac
		{ \sum_{k}  \,
		[R(  \frac{ K_{k} }{s} ,\Omega_{k})]_{i} }
		{J_0 \,  \sum_{k} \, [R(\mathds{1},\Omega_{k})]_{i}}
\label{model_best}
\end{equation}
where we are following the notation in Equations~(\ref{3D_update_MC}) and  (\ref{3D_update}).
$\Omega_k$ represents the orientation of the $k^{th}$ diffraction pattern.
This resembles Equations~(\ref{3D_update_MC}) and  (\ref{3D_update})
if the orientational probability was calculated as zero everywhere but the correct orientation.

\subsubsection{Best Possible Variable Fluence Reconstruction}
In the variable flunce case,
we need the $k^{th}$ diffractions patterns' associated orientation and fluence,  $\Omega_k$ and  $J_k$ respectively,
to construct a merge in the same manner as Equation~(\ref{3D_update_MC_J}).
The best possible merge in the variable fluence case can be defined as follows
\begin{equation}
	M^{best}_{i}=\frac
		{ \sum_{k}  \,
		[R(  \frac{ K_{k} }{s} ,\Omega_{k})]_{i} }
		{ \,  \sum_{k} J_k \, [R(\mathds{1},\Omega_{k})]_{i}}
\end{equation}

\subsubsection{Best Possible Two State Reconstruction}

To construct the best possible merge in the two-state case,
we need to insert the $k^{th}$ diffraction pattern at the appropriate orientation into the correct model.
We write this as
\begin{equation}
	M^{best}_{\sigma i}=\frac
		{ \sum_{k}  \, \delta_{\sigma \sigma_k}
		[R(  \frac{ K_{k} }{s} ,\Omega_{k})]_{i} }
		{J_0 \,  \sum_{k} \, \delta_{\sigma \sigma_k}[R(\mathds{1},\Omega_{k})]_{i}} \;.
\end{equation}
Following the notation of equation \ref{3D_update_MC_2}. $\sigma_k$ represents the state of the $k^{th}$ diffraction pattern.

\subsubsection{Error Measure}
An error measure was scripted to judge the accuracy of a reconstruction, $M$, compared to the appropriate $M^{best}$.
The reconstruction takes a random orientation relative to $M^{best}$,
so we employ an exhaustive approach which calculates the error over a set of orientations, $\{ \Omega_r \}$, and taking the minimum value
 \begin{equation}
	\mathrm{Error} \equiv 
		\mathrm{min} \left( \left\{ 
		\sqrt{ \frac
		{\sum_i ( M_i - [ R( M^{best}, \Omega_r ) ]_i \, ) ^ 2 }
		{ \sum_i { M^{best}_i }^ 2} 
		} \right\} \right) \;.
\end{equation}
The operator $R$ here rotates $M^{best}$ by $\Omega_r$.
The curly brackets represent the set of values resulting from performing the calculation for all of the orientations in $\{ \Omega_r \}$.
Performing this operation on the models for each iteration yields a curve we can plot to display how the error improves with iterations.

\subsubsection{Differences Measure}

 Also plotted are relative differences between models over iterations which give an idea of the convergence of the algorithm.
 As the reconstruction is reaching a steady state, the difference between consecutive models will drop to near zero.
 The difference measure is defined as follows
 \begin{equation}
	\Delta M^t \equiv \frac{
	\sqrt{
	\sum_i ( M^t_i - M^{ t - 1 }_i  ) ^ 2
	}}
	{
	\sum_i \frac{1}{2} ( M^t_i + M^{t - 1}_i )
	} \;.
\end{equation}

 \section{Results}
 
 \subsection{Comparison Results}
 
\begin{figure}
\centering
\includegraphics[width=\linewidth]{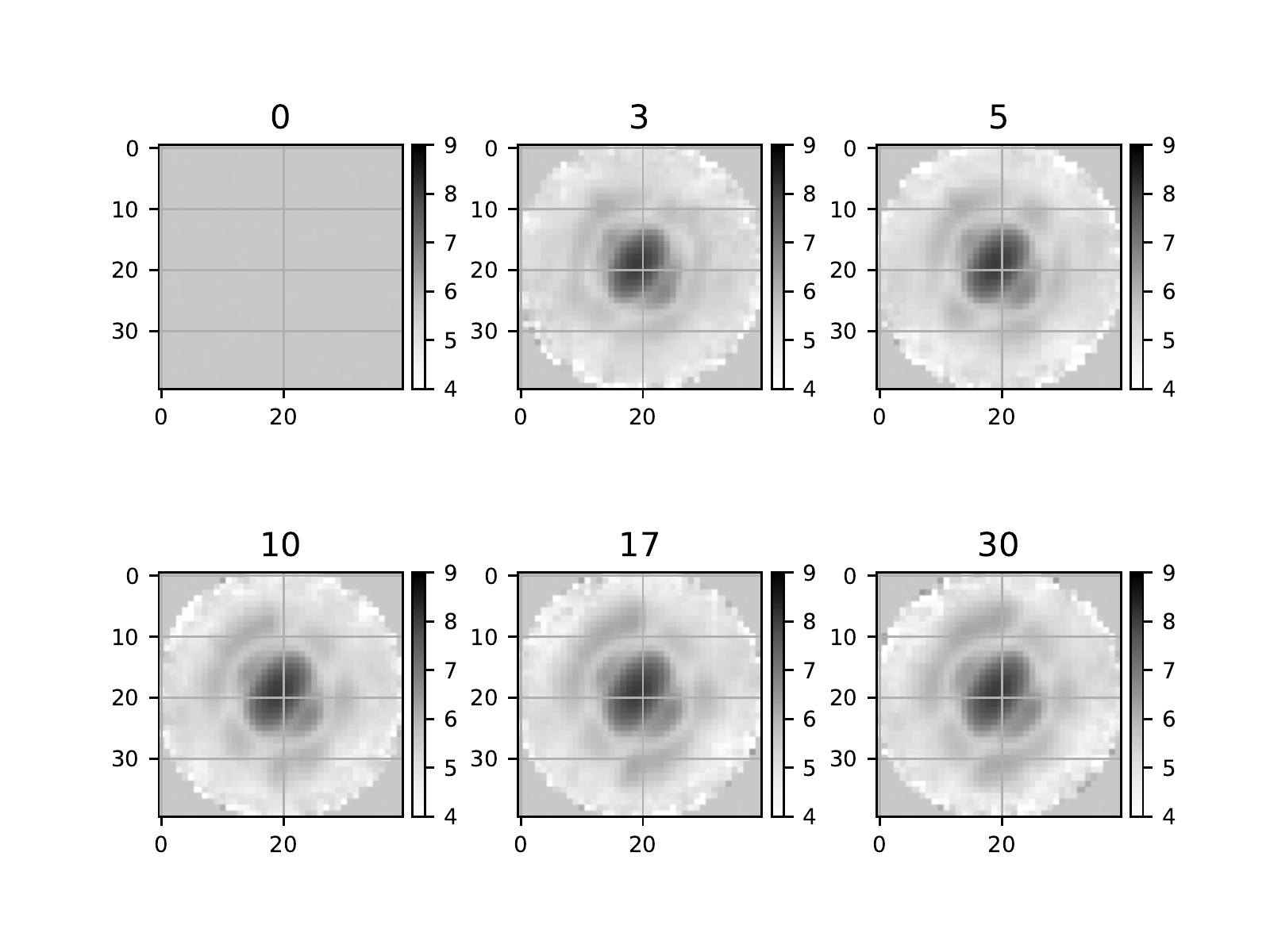}
\caption{
Pictured is the progression of the Monte Carlo EMC reconstruction.
Equatorial slices of the 3D diffraction volume are plotted on a log scale at sample points along the reconstruction.
The number labels the number of iterations of the EMC loop.
}
\label{mc_progress}
\end{figure}

\begin{figure}
\centering
\includegraphics[width=\linewidth]{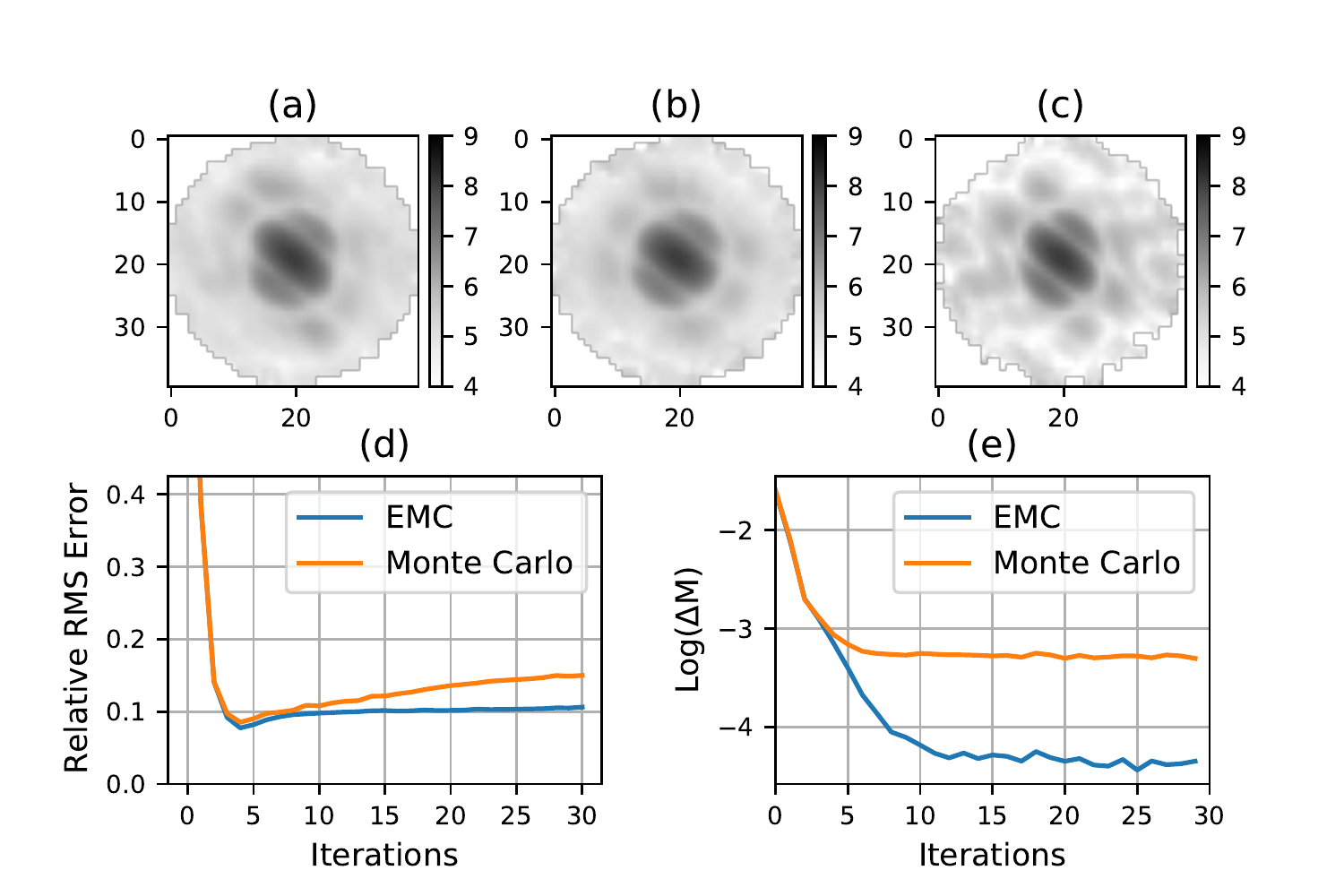}
\caption{
Pictured is a comparison of the results of the traditional EMC algorithm and the Monte Carlo implementation.
Shown in the top row are equatorial slices of the 3D diffraction volumes plotted on a log scale. 
We have
(a) the final model of the Monte Carlo reconstruction,
(b) the final model of traditional EMC reconstruction,
(c) the best possible reconstruction,
(d) error over iterations for both methods, and
(e) log of the model difference metric for both methods.
}
\label{all_three}
\end{figure}

The regular sampling and Monte Carlo algorithms were run on the same data set of 4000 simulated diffraction patterns
described in section \ref{sec_constant_fluence_data} and the same random input model.
The Metropolis walks were performed for 800 steps, 
the first 200 of them were not considered in the average as a burn-in period to allow for the walk to become uncorrelated with its starting position.
The regular sampling algorithm was performed with a set of 420 orientations.
Progression of the Monte Carlo reconstruction is pictured in figure \ref{mc_progress},
and results of the traditional and Monte Carlo implementations are shown in figure \ref{all_three}.
The Monte Carlo reconstruction is of comparable quality to the regular sampling reconstruction,
but is likely limited by the noise floor inherent in the Monte Carlo technique,
as seen in the asymptotic behavior in plots in figure \ref{all_three}b and \ref{all_three}c.
Though neither algorithm run here is optimized,
the Monte Carlo reconstruction showed the encouraging result that the regular sampling algorithm took 27 hours whereas the Monte Carlo took only 25 hours.

\subsection{Variable Fluence Result}
\begin{figure}
	\centering	
	\includegraphics[width=\linewidth]{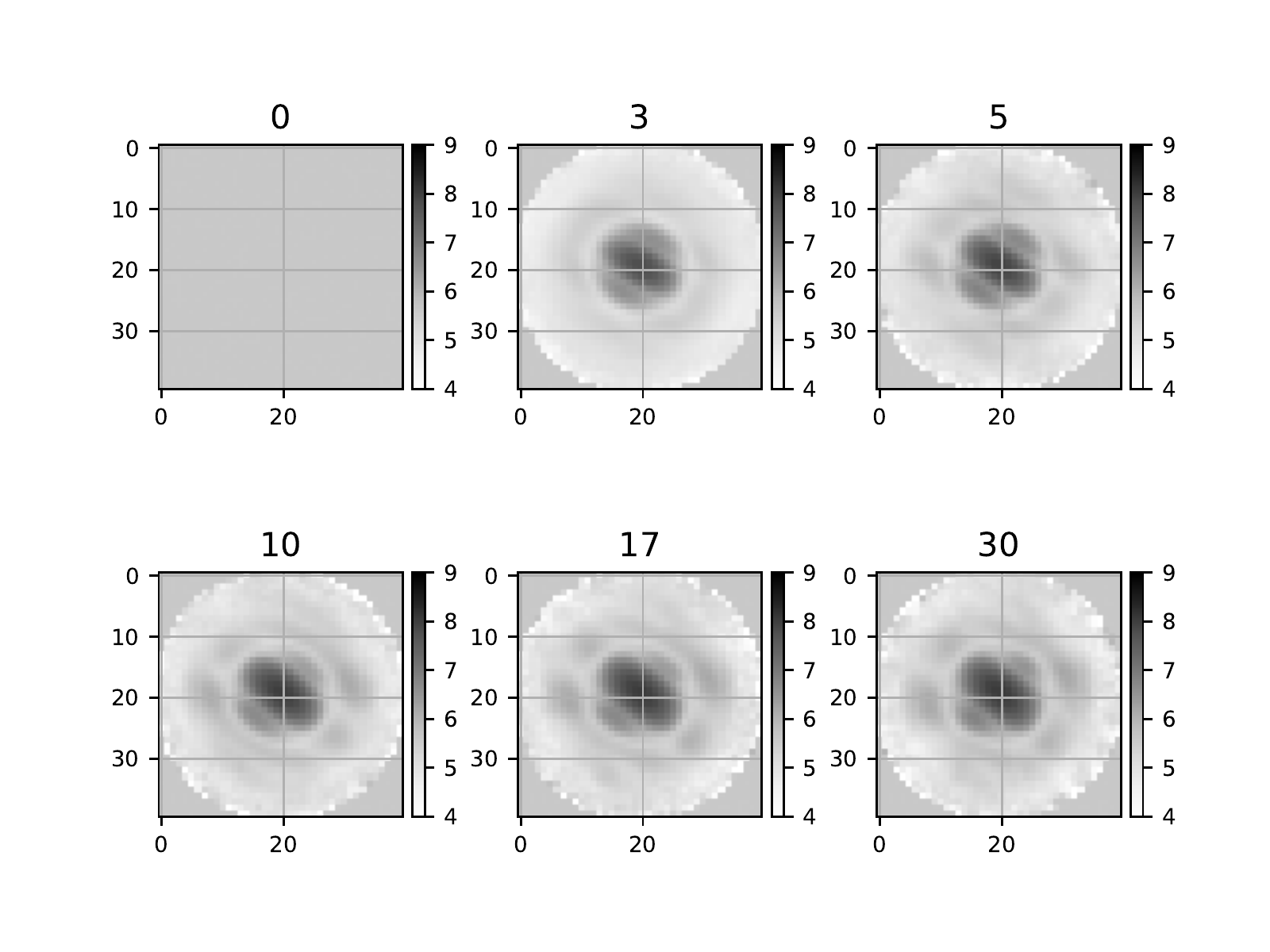}
	\caption{
Progression of the variable fluence Monte Carlo EMC reconstruction. Shown are
equatorial slices of the 3D diffraction volume on a log scale at sample points along the reconstruction.
The number labels the number of iterations of the EMC loop.
	}
	\label{variable_fluence_progress}
\end{figure}

\begin{figure}
	\centering	
	\includegraphics[width=\linewidth]{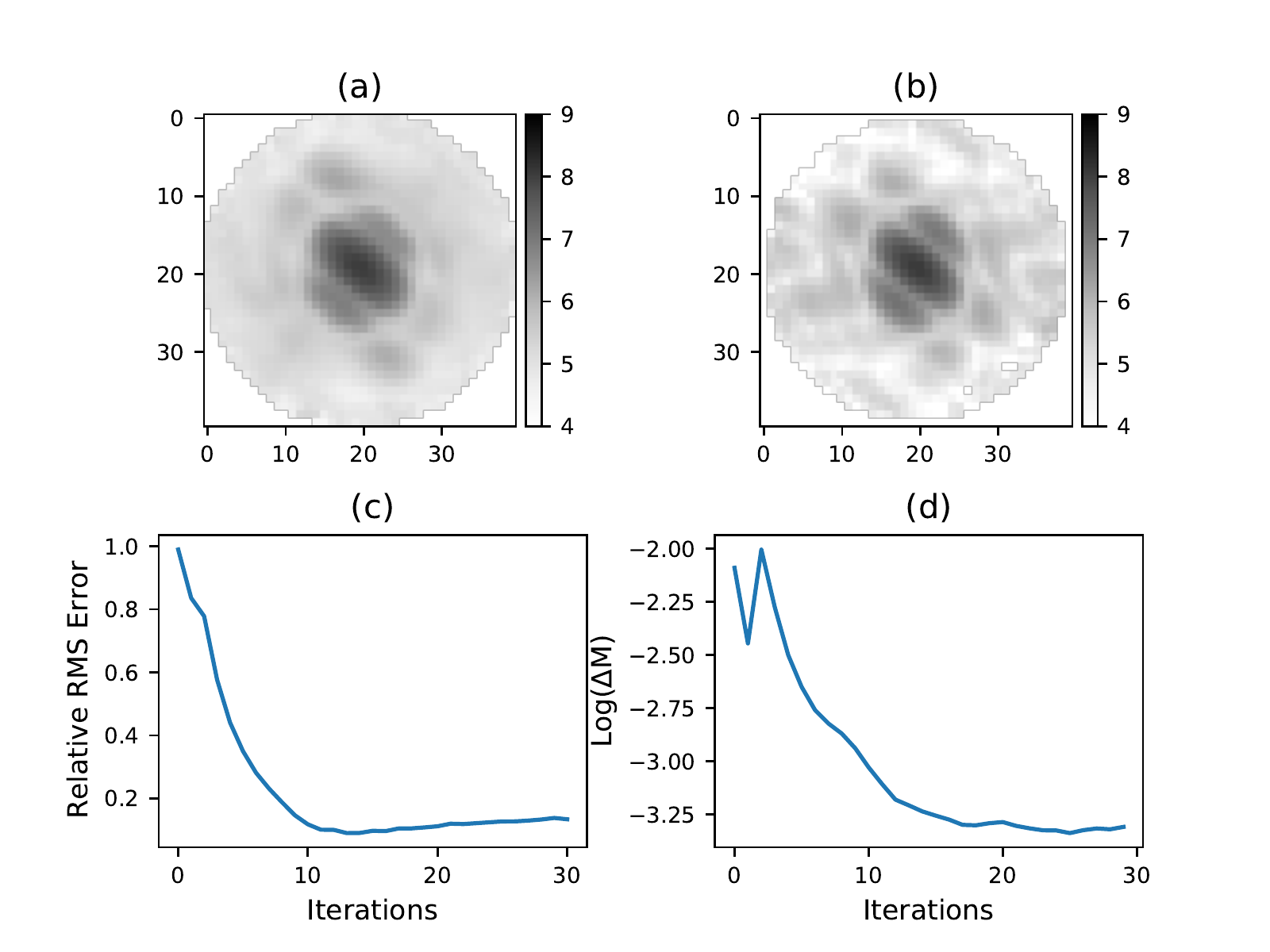}
	\caption{
Shown are the results of the variable fluence Monte Carlo reconstruction.
Pictured in the top row are equatorial slices of the 3D diffraction volumes shown on a log scale.
We have
(a) the final model of the variable fluence Monte Carlo reconstruction,
(b) the best possible reconstruction,
(c) error over iterations, and
(d) the log difference metric over iterations.
	}
	\label{variable_fluence}
\end{figure}

The variable fluence algorithm was run on 4000 simulated diffraction patterns described in section \ref{sec_variable_fluence_data}.
Each Metropolis walk was performed for 800 steps, the first 200 of which were discarded as burn-in.
Progress along the reconstruction are shown in figure \ref{variable_fluence_progress},
and results are shown in figure \ref{variable_fluence}.
The reconstruction reaches a similar level of error and convergence time as the constant fluence Monte Carlo EMC algorithm.
The error measure over iterations, figure \ref{variable_fluence}c,
shows a phenomenon of slightly increasing as the reconstruction is approaching the steady state,
but this is not entirely surprising because the reconstruction is not minimizing this bulk error measurement,
and there may be blurrier intermediate states that have a better error according to this particular metric than the sharper final state.

\subsection{Two-State Result}
\begin{figure}
	\centering	
	\includegraphics[width=\linewidth]{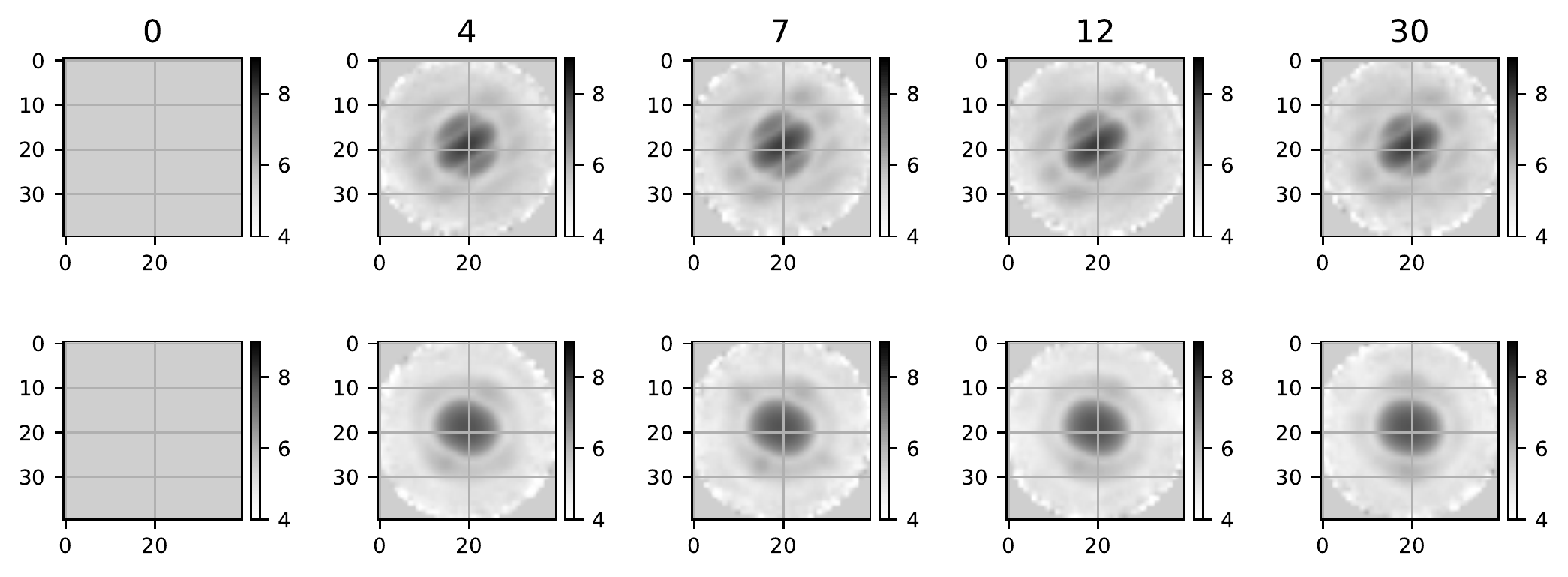}
	\caption{
Progression of the two-state Monte Carlo EMC reconstruction.
Log scale equatorial slices of the two diffraction volumes are paired vertically,
and each column shows them at sample points sample points along the reconstruction.
The number labels the number of iterations of the EMC loop.
	}
	\label{two_state_progress}
\end{figure}

\begin{figure}
	\centering	
	\includegraphics[width=0.9\linewidth]{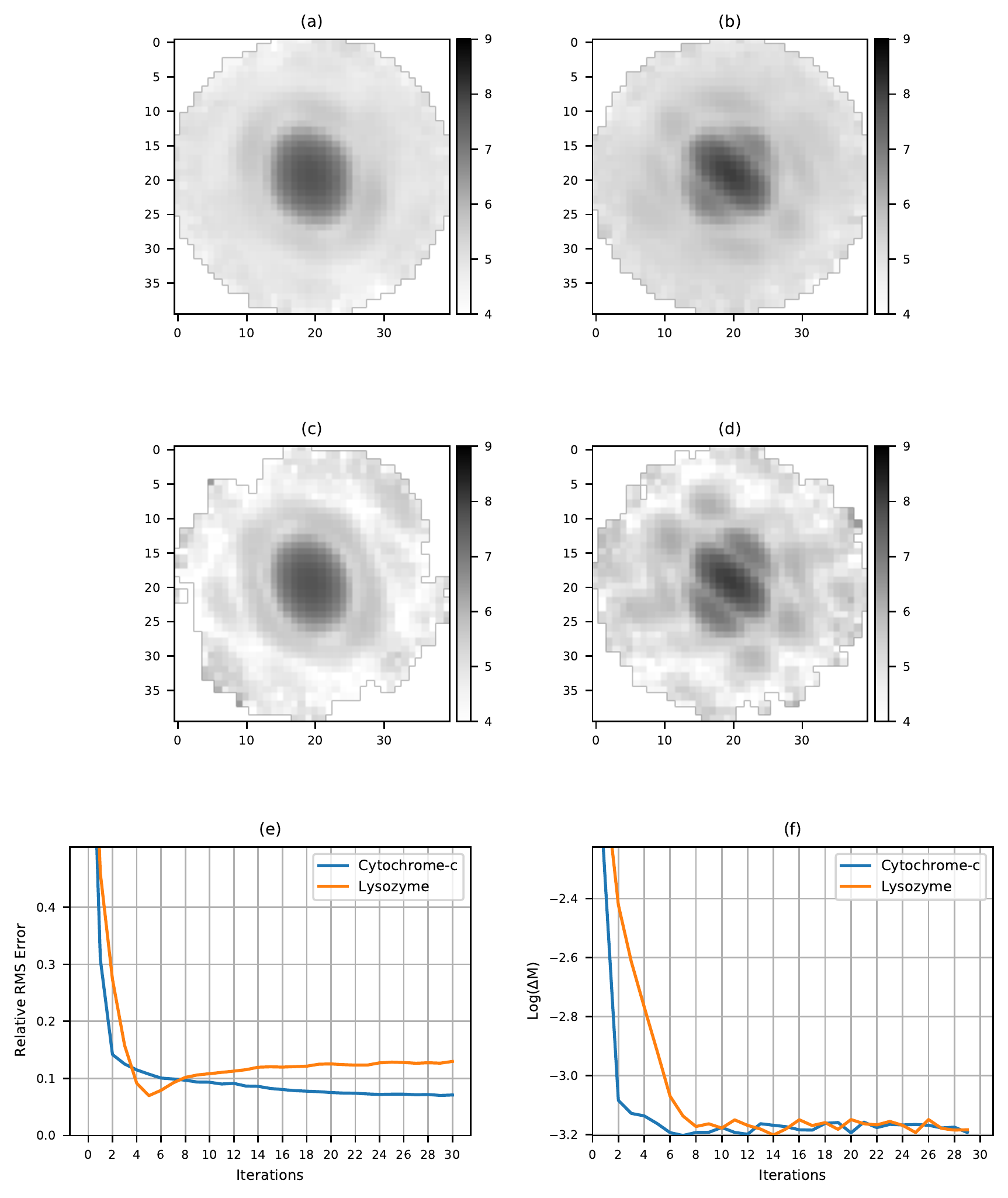}
	\caption{
Shown are the results of the two-state Monte Carlo EMC reconstruction.
Pictured in the top four plots are equatorial slices of the 3D diffraction volumes shown on a log scale.
We have
(a) the final model of the variable fluence Monte Carlo reconstruction of lysozyme,
(b) the final model of the variable fluence Monte Carlo reconstruction of cytochrome-c,
(c) the best possible reconstruction of lysozyme,
(d) the best possible reconstruction of cytochrome-c,
(e) error over iterations of both diffraction volumes, and
(f) the log difference metric over iterations.
	}
	\label{two_state}
\end{figure}

The two-state algorithm was run on 4000 simulated diffraction patterns described in section \ref{sec_two_state_data}
of which 40\% of the diffraction patterns are from lysozyme and 60\% are from cytochrome-c.
Each Metropolis walk was performed for 800 steps, the first 200 of which were discarded as burn-in.
Progress along the reconstruction is pictured in figure \ref{two_state_progress},
and the results are shown in figure \ref{two_state}.

\section{Conclusion}
The Monte Carlo Expand Maximize Compress algorithm has shown initial success in the regimes presented in this paper,
and can be considered a valid alternative implementation of the EMC algorithm.
A simple benefit of this implementation is that is does not require a discretization of orientation space
or of any additional parameters that may be part of a desired application
and thus can be rapidly adapted to other regimes without having to redesign the grid over which one would be performing numerical integration.
For instance, we can readily adapt this algorithm to the case of crystal diffraction data with Bragg peaks
and have the algorithm explore the tight space of orientations around the peaks instead of all orientations with only minimal changes.
The main benefit remaining to be properly demonstrated is that,
in the case of experiments with unknown degrees of freedom in addition to orientation and fluence such as
target mixture, target conformation, wavefront jitter, or crystal parameters,
the Monte Carlo technique should be able to more efficiently handle these higher dimensional regimes,
as is usually the case for Monte Carlo compared to deterministic, exhaustive approaches.

\appendix
\section{Insertion function}
\label{Appen_A_trilinear_insertion}
The insertion of the 2D diffraction pattern into a 3D array is handled by a function we call $R(X,\Omega)$ defined as
\begin{equation}
	R(X,\Omega) = \sum_{j} \text{TrilinearInsert}(X_j, \mathcal{R}(\Omega) \cdotp \vec{q}_{j} ) \;,
\end{equation}
where the first input $X$ is a vector that has an entry for each pixel of the detector, indexed by $j$.
The second input $\Omega$ is the orientation at which the pixels are to be inserted.
The output is a vector whose entries are the voxel values of a 3D array, indexed by $i$.
We will call the $j^{th}$ q-vector of the detector Ewald sphere $\vec{q}_{j}$ and $\mathcal{R}(\Omega)$ will represent the rotation matrix associated with $\Omega$.
The dot is to be understood as matrix multiplication. The product $\mathcal{R}(\Omega) \cdotp \vec{q}_{j}$ will result in the detector pixel q-values being rotated to the desired orientation.
The function performs a trilinear insertion for each value of $X$ and each corresponding oriented detector pixel q-vector.

The trilinear insert function takes a value and a q-vector as inputs and outputs a 3D array that is zero everywhere
except at the eight voxels whose q-vectors define the cube that encompasses the input q-vector point. The insertion into the $i^{th}$ voxel can be written as
\\
\begin{equation}
	\Big[\text{TrilinearInsert}( v, \vec{q} )\Big]_i = v \xi_i(  \vec{ q } ) \;,
\end{equation}
\\
where the function $\xi_i(  \vec{ q } )$ is defined as

\begin{equation}
	\xi_i(  \vec{ q } ) =  
	\begin{cases}
	\left( 1 - \frac{| q_x - q'_{ix} |}{a} \right) 
	\left( 1 - \frac{| q_y - q'_{iy} |}{a} \right)
	\left( 1 - \frac{| q_z - q'_{iz} |}{a} \right)  &   
	\text{ if } | q_x - q'_{ix} | \leq a 
	\text{ and  } | q_y - q'_{iy} | \leq a 
	\text{ and }  | q_z - q'_{iz} | \leq a 
	\\
	0
	&  \text{ otherwise, }
	\end{cases}
\label{eq_trilinearInsertionXi}
\end{equation}
where ${\vec{q}}'_i$ is the position of the $i^{th}$ voxel and $a$ is the grid spacing between the voxels in q-space.
The piecewise construction of Equation~(\ref{eq_trilinearInsertionXi}) ensures $\xi_i$ is nonzero only in the cube centered around ${\vec{q}}'_i$ with edge length $2a$.

\section{Interpolation function}
\label{Appen_B_trilinear_interp}
The interpolation of the 3D model along an Ewald slice at an arbitrary orientation is handled by the function $E(M,\Omega)$.
This function takes a 3D array of numbers, $M$, and an orientation, $\Omega$, as inputs and outputs a vector with an entry for each detector pixel indexed by $j$.

\begin{equation}
	E_{j}(M,\Omega) = \text{TrilinearInterpolate}( M,  \mathcal{R}(\Omega) \cdotp \vec{q}_{j} )
\end{equation}
As in Appendix~\ref{Appen_A_trilinear_insertion}, the symbol $\mathcal{R}$ represents a rotation matrix, and the dot is to be understood as matrix multiplication.

The trilinear interpolation works by finding the eight voxel grid positions defining a cube surrounding the position at which we wish to interpolate,
evaluating the function at those points,
and finding a weighted average of those values with weights determined by the position of interest relative to the surrounding points. 

The weights are determined by dividing the cube into eight cells by drawing the three planes parallel to the grid intersecting at the point of interest.
The volume of the cell across from a particular corner serves as the weight of that corner's evaluation in the weighted average.

It is interesting to note that the functions $E$ and the function $R$ from Appendix~\ref{Appen_A_trilinear_insertion} are not inverses of each other.
$E( R( X )) = X$ but $R( E( M)) \neq M $

\ack{We thank Prof. John C. H. Spence for his support, guidance, and suggestions. This work was supported by NSF awards 1817862, 1231306, and 1943448 }

\referencelist[ref]

\end{document}